\documentstyle[11pt,amssymb]{article}   
\def\AFOUR{%
\setlength{\textheight}{9.0in}%
\setlength{\textwidth}{5.75in}%
\setlength{\topmargin}{-0.375in}%
\hoffset=-.5in%
\renewcommand{\baselinestretch}{1.17}%
\setlength{\parskip}{6pt plus 2pt}%
}
\AFOUR                                           
\def\car{\mathop{\square}}
\def\carre#1#2{\raise 2pt\hbox{$\scriptstyle #1$}\car_{#2}}

\makeatletter
\def\section{\@startsection {section}{1}{\z@}{-3.5ex plus -1ex minus
 -.2ex}{2.3ex plus .2ex}{\large\bf}}
\def\subsection{\@startsection{subsection}{2}{\z@}{-3.25ex plus -1ex minus
 -.2ex}{1.5ex plus .2ex}{\normalsize\bf}}
\makeatother
\makeatletter
\@addtoreset{equation}{section}

\makeatother
\newcommand{\nc}{\newcommand}
\newcommand{\rnc}{\renewcommand}
\nc{\be}{\begin{equation}}
\nc{\ee}{\end{equation}}
\nc{\bea}{\begin{eqnarray}}
\nc{\eea}{\end{eqnarray}}
\def\slash#1{\setbox0=\hbox{$#1$}#1\hskip-\wd0\hbox to\wd0{\hss\sl/\/\hss}}

\def\href#1#2{{#2}}

\rnc{\a}{\alpha}
\nc{\ab}{\bar{\a}}
\nc{\ap}{\a^{+}}
\nc{\abm}{\ab^{-}}
\rnc{\b}{\beta}
\nc{\bb}{\bar{\b}}
\nc{\bbp}{\bb_{\zb}^{+}}
\nc{\bm}{\b_{z}^{-}}
\nc{\oa}{\overline{\a}}
\nc{\ob}{\overline{\b}}
\rnc{\gg}{\gamma}
\rnc{\d}{\delta}
\nc{\f}{\phi}
\nc{\fb}{\bar{\phi}}
\nc{\vf}{\varphi}
\nc{\p}{\psi}

\rnc{\c}{\chi}
\nc{\la}{\lambda}
\nc{\m}{{\mathrm m}}
\nc{\n}{\nu}
\rnc{\o}{\omega}
\nc{\Om}{\Omega}
\rnc{\t}{\theta}
\nc{\eps}{\epsilon}
\rnc{\S}{\Sigma}
\nc{\F}{\Phi}
\nc{\trac}[2]{{\textstyle\frac{#1}{#2}}}
\nc{\ex}[1]{\mbox{e}^{\,\textstyle#1}}
\nc{\mat}[4]{\left(\begin{array}{cc}#1&#2\\#3&#4\end{array}\right)}
\nc{\som}[9]{\left(\begin{array}{ccc}#1&#2&#3\\#4&#5&#6\\#7&#8&#9%
\end{array}\right)}
\nc{\tr}{\mathop{\mbox{tr}}\nolimits}
\nc{\ad}{\mathop{\mbox{ad}}\nolimits}
\nc{\Tr}{\mathop{\mbox{Tr}}\nolimits}
\nc{\Det}{\mathop{\mbox{Det}}\nolimits}
\nc{\rk}{\mathop{\mbox{rk}}\nolimits}
\nc{\ra}{\rightarrow}
\nc{\Ra}{\Rightarrow}
\nc{\LRa}{\Leftrightarrow}
\nc{\ot}{\otimes}
\rnc{\ss}{\subset}
\nc{\nul}{\noindent\underline}
\nc{\non}{\nonumber\\}
\nc{\subs}[1]{{\vspace*{0.5cm}}%
{\noindent\underline{#1}}{\addcontentsline{toc}{subsection}{#1}}%
{\vspace*{0.3cm}}}
\nc{\zb}{\bar{z}}
\rnc{\lg}{\frak{g}}
\nc{\lt}{\frak{t}}
\nc{\lk}{\frak{k}}
\nc{\lh}{\frak{h}}
\nc{\pik}{\Pi_{\lk}}
\nc{\pip}{\Pi_{+}}
\nc{\pim}{\Pi_{-}}
\nc{\pih}{\Pi_{\lh}}
\nc{\jz}{J_{z}}
\nc{\jzh}{\jz^{\lh}}
\nc{\jzp}{\jz^{+}}
\nc{\jzm}{\jz^{-}}
\nc{\del}{\partial}
\nc{\dz}{\del_{z}}
\nc{\dzb}{\del_{\bar{z}}}
\nc{\az}{A_{z}}
\nc{\azb}{A_{\bar{z}}}
\nc{\g}{g^{-1}}
\nc{\dw}{\Delta_{W}}
\nc{\Ad}{{\mbox{Ad}}}
\nc{\ks}{Ka\-za\-ma-\-Su\-zu\-ki}
\nc{\KS}{\ks}
\nc{\ksm}{\ks\ model}
\rnc{\AA}{{\Bbb A}}
\nc{\BB}{{\Bbb B}}
\nc{\CC}{{\Bbb C}}
\nc{\PP}{{\Bbb P}}
\nc{\cpm}{\CC\PP(m)}
\nc{\cpn}{\CC\PP(n)}
\nc{\cp}[1]{\CC\PP(#1)}
\nc{\gmn}{G(m,m+n)}
\nc{\gmnk}{\gmn_{k}}
\nc{\cO}{{\cal O}}
\nc{\bcO}{\bar{\cO}}
\nc{\bO}{\bar{O}}
\nc{\oQ}{\overline{Q}}
\nc{\ie}{{\it i.e.~}}
\nc{\eg}{{\it e.g.~}}
\begin{document}
\global\parskip=4pt
%%%%%%%%% title page %%%%%%%%%%%%%%%%%%%%%%%%%%%%%%%%%%%%%%%%
\makeatother\begin{titlepage}
\begin{flushright}
{SISSA/128/99/EP}
\end{flushright}
\vspace*{0.5in}
\begin{center}
{\Large\bf Fivebrane instantons and higher derivative}\\ 
{\Large\bf couplings in type I theory}\\
\vskip .3in
\makeatletter
\begin{tabular}{c}
{\bf Amine B. Hammou}, \footnotemark 
\\ 
SISSA, Trieste, Italy, \\
\end{tabular}

\begin{tabular}{c}
{\bf Jose F. Morales}, \footnotemark
\\ 
INFN, Sezione di ``Tor Vergata'', Roma, Italy, \\
\end{tabular}

\end{center}
\addtocounter{footnote}{-1}%
\footnotetext{e-mail: amine@sissa.it}
\addtocounter{footnote}{1}%
\footnotetext{e-mail: morales@roma2.infn.it}

\vskip .80in
\begin{abstract}
\noindent 
We express the infinite sum of D-fivebrane instanton corrections 
to ${\cal R}^2$ couplings in ${\cal N}=4$ type I string vacua, in terms
of an elliptic index counting $\frac{1}{2}$-BPS excitations in the 
effective $Sp(N)$ brane theory. We compute the index explicitly in the
infrared, where the effective theory is argued to flow to an orbifold CFT. 
The form of the instanton sum agrees completely with the 
predicted formula from a dual one-loop computation in  
type IIA theory on $K3\times T^2$. 
The proposed CFT provides a proper description of the whole spectrum
of masses, charges and multiplicities for $\frac{1}{2}$- and
$\frac{1}{4}$- BPS states, associated to bound states of D5-branes
and KK momenta. These results are applied to show how  
fivebrane instanton sums, entering
higher derivative couplings which are sensitive to $\frac{1}{4}$-BPS 
contributions, also match the perturbative results in the
dual type IIA theory.  

\end{abstract}

%\vskip .2in
\vfill
\noindent
{\it PACS: 11.25.-w, 11.25.Hf, 11.25.Sq}

\noindent
{\it Keywords: Instantons, D5-brane, Thresholds, BPS, CFT}

\makeatother
\end{titlepage}
%%%%%%%%%% end of title page %%%%%%%%%%%%%%%%%%%%%%%%%%%%%%
\begin{small}
%\tableofcontents
\end{small}

\setcounter{footnote}{0}

\section{Introduction}
           
String dualities often map worldsheet instanton effects on
one string vacuum to spacetime instantons in the dual one 
\cite{witten1,seeHM}.
A perturbative computation in one side can then be used to shed light
about the general rules of the instanton calculus in string theory.
This idea motivates the study of \cite{bfkov}, where 
heterotic world-sheet instanton corrections to four-derivative couplings 
in toroidal compactifications to eight dimensions were translated 
into a sum of D-instanton effects  
in the dual type I side. The analysis was extended to lower dimensional
($D> 4$) hetetoric/type I toroidal compactifications 
\cite{thresholdI}
and to ${\cal N}=4$ dual pairs of type II string vacua \cite{bgmn,ghmn}. 
In each of the cases the 
N-instanton corrections, associated to bound states of N D-string 
wrapping the $T^2$-spacetime torus, were expressed in terms of elliptic
genera counting the number of $\frac{1}{2}$-BPS excitations in the 
relevant $O(N)$ gauge theories. The analyses were always restricted to
$D > 4$ dimensions where the less understood fivebrane physics  
is irrelevant. The aim of this paper is to extend to $D=4$ dimensions
these results by explicitly computing the D5-instanton effects in the
type I brane theory.

We will consider instanton corrections to higher derivative 
couplings in toroidally compactified type I theory down to four
dimensions. The only sources
of non-perturbative corrections (to the kind of couplings we will
consider) in these string vacua
are associated to Euclidean 
D5-branes wrapping entirely the $T^6$-torus, whose BPS excitations
can be properly described, as we will 
see, by an ${\cal N}=(4,4)$
orbifold CFT. 
Less supersymmetric instanton configurations would
have too many fermionic zero modes to be soaked by the  
vertex insertions (typically ${\cal R}^2$), 
while D-string instantons leads to ${\cal N}=(8,0)$
effective sigma models \cite{gmnt},
where the vertex insertions  can soak 
at most four of the eight left moving fermionic zero modes. 

We will work out the details for ${\cal R}^2$ couplings and comment 
about the generalizations to other four
and higer derivative ``BPS'' saturated terms. These couplings   
are special in that they receive corrections only from states 
saturing the BPS bounds and they have been extensively 
studied in many contexts \cite{hm}-\cite{ls}.
In this paper we will concentrate in the non-perturbative part
of the corrections to $D=4$ effective lagrangians in type I
vacua with sixteen supercharges. 
The instanton sums will be always expressed 
in terms of an elliptic genus in the effective
$Sp(N)$ gauge theory, which encodes the information about
masses, charges and multiplicties of $\frac{1}{2}$ BPS
excitations in the corresponding D5-brane system. 
We will compute the ``quasi'' topological index and show that the 
form of the instanton corrections to the associated couplings 
agree with the predictions from duality to type IIA
on $K3\times T^2$. 
Interestingly, the CFT description of the infrared 
limit of the D5-brane world-volume theory reproduces the
right multiplicities even for $\frac{1}{4}$-BPS states, associated to
D5-KK bound states.
We apply this result to show that instanton corrections
to certain higher derivative couplings, sensitive to these states,
agree again with the 
fundamental type IIA results\footnote{The couplings under consideration, 
are closely related to those studied in \cite{ls}.}.

The paper is organized as follows. In section 2 we compute D5-instanton
corrections to ${\cal R}^2$-couplings in type I theory on $T^6$, 
and compare it to the perturbative result obtained in 
the dual type IIA theory.
In section 3, we briefly discuss the extension of such results 
to other higher derivative terms, which are sensitive to $\frac{1}{4}$
BPS contributions.
Section 4 summarizes the 
results and present some conclusions.

\section{${\cal R}^2$ couplings  in type IIA/type I 
${\cal N}=4$  string vacua}  

As we have discussed in the introduction, worldsheet/spacetime instantons
in type IIA theory on $K3\times T^2$/type I theory on $T^6$
are mapped to each other under the triality map
\be
T_{IIA}=S_{H}=S_{I}.\label{dualitymap}
\ee
The subscripts $IIA,H,I$ refer to the type II, Heterotic and type I
theories, while the complex moduli\footnote{In this paper, 
all quantities having dimension of a length will be 
understood in units of $2\pi \sqrt{\alpha^\prime}$.} 
\bea 
T_{IIA}&=&B_{45}+i\sqrt{G}\nonumber\\
S_H&=&a+i\frac{{\cal V}_6}{\lambda^2}\nonumber\\
S_I&=&
a+i\frac{{\cal V}_6}{\lambda}\label{map}.
\eea
describe the complexified Kahler structure of the $T^2$ torus 
in the type IIA side, 
and the four dimensional
complexified string couplings  in the heterotic and type I string vacua
respectively with ${\cal V}_6$ the  $T^6$-volume. 
More precisely, $a={\tilde B}_{456789}$, where $\tilde B$ 
is defined by $d{\tilde  B}= ^* d B$, $B$ being
the second rank, antisymmetric tensor in the 
corresponding string theory, and $\lambda=e^{\phi}$
the ten-dimensional string coupling.

In this section we will show how the perturbative,
world-sheet instanton contribution to the ${\cal R}^2$
coupling in the type IIA theory, can be directly
reproduced in type I theory, using the
effective six dimensional, ${\cal N}=1$, 
gauge theory  of D5 branes, in the limit
where it flows to a two dimensional ${\cal N}=(4,4)$
orbifold Conformal Field Theory, after dimensional
reduction on a four-torus.

\subsection{One-loop ${\cal R}^2$ couplings in type IIA}

Let us start by recalling the results for 
the moduli dependence $\Delta_{\rm gr}(T,U)$ of ${\cal R}^2$-couplings
in type IIA theory on $K3\times T^2$ \cite{hm,ko}
\footnote{We will follow 
the notations and normalizations of \cite{ko}.}.
Perturbative corrections to ${\cal R}^2$ terms in $(2,2)$ string vacua 
are expected to arise only at one-loop level, and depend only on the
$T$($U$) complex modulus describing the Kahler(complex) structure
of the $T^2$ torus in the type IIA(IIB) string compactifications. 
The type IIA result can then be written as \cite{hm,ko} 
\be
\partial_T \Delta_{\rm gr} (T) = 
\int_{\cal F} {d^2 \tau\over\tau_2} \partial_T B_4\, ,
\label{delta}
\ee
where $B_4$, is an index counting the number of $\frac{1}{2}$-BPS 
string states (the helicity supertrace). The index can be defined
as \cite{ko}
\be
B_4= \left\langle \left(\lambda_{\rm L}+
\lambda_{\rm R}\right)^4 \right\rangle
= \left({4 \atop 2}\right)\frac{1}{(2\pi i)^2}\left.\partial_v^2 \chi(v|\tau)
\right|_{v=0}\equiv
{  6\over 16 \pi^4} \left.\partial_v^2 \partial_{\bar{v}}^2  Z(v,\bar{v})
\right|_{v = \bar{v} =0}\, ,
\label{b4} 
\ee
in terms of the helicity supertrace generating function 
\be
Z(v,\bar v)=\Tr^\prime
q^{L_0-{c\over 24}}\,
\bar q^{\bar L_0 - {\bar c \over 24}}\,
e^{2\pi i\left(v\lambda_{\rm L}-\bar v \lambda_{\rm R}\right)}
\, .
\label{Zvv}
\ee
$\lambda_{L(R)}$ being the left(right) mouving helicity operators, and
the prime on $\Tr$ means the omission of 
the space-time bosonic zero modes 
contributions, and $q\equiv e^{2\pi i \tau}$.
One can see that $Z(v,\bar{v})$ receives 
in general contributions from both BPS and non-BPS string states. 
This is not the case for the chiral supertrace $\chi(v|\tau)$,
introduced in (\ref{b4}). 
Indeed, the insertion of two $\lambda_R$'s in (\ref{Zvv}) soaks
precisely four right-moving fermionic zero modes 
(after spin structure sums), while massive bosonic
and fermionic right moving excitations, cancel against each other by 
supersymmetry, giving as a result the holomorphic function
$\chi(v|\tau)$\footnote{The chiral trace $\chi(v|\tau)$ can be considered
as the generating function for the asymmetric supertraces introduced
in \cite{ls}.}. This chiral supertrace encodes all the
information about BPS multiplicities and charges, and we will
refer to it as the ``elliptic genus'' of the corresponding
conformal field theory.   
Specializing to the case of type IIA string theory on $K3 \times T^2$,
we get, after summing over
the spin structures,  
\be
Z(v,\bar v)=8\left|\xi(v)\frac{\vartheta_1^2(\frac{v}{2})}{\eta^6}\right|^2
\sum_{i=1}^4 \left|\frac{\vartheta_i^2(\frac{v}{2})}{\vartheta^2_i(0)}\right|^2
\Gamma_{2,2}
\label{zvvII}
\ee
where
\be
\xi(v)\equiv \frac{sin \pi v}{\pi}\frac{\theta_1^\prime(0)}{\theta_1(v)}
\ee
is the contribution of the spacetime boson coupled to the helicity.
Substituting (\ref{zvvII}) in (\ref{b4}), one finds \cite{ko}
\be
B_4 = 36~\Gamma_{2,2}.
\label{b4IIa}
\ee
The $\Gamma_{2,2}$ lattice sum can be written as  
\be
\Gamma_{2,2}=
\frac{T_2}{\tau_2}\,
 \sum_{M\in GL(2,Z)}
 e^{ 2\pi i  T {\rm det}M }
e^{- \frac{\pi T_2 }{ \tau_2 U_2 }
\big| (1\; U)M  \big( {\tau \atop -1} \big) \big| ^2},
\label{lattice}
\ee
where the sum runs over all possible world-sheet instantons
\bea
\left( {X^4 \atop X^5} \right)=M\,\left( {\sigma^1 \atop \sigma^2} \right)
\equiv \pmatrix {m_1 & n_1 \cr m_2 & n_2}
\left({\sigma^1 \atop \sigma^2} \right),   
\eea
with worldsheet and target space coordinates $\sigma^1$, $\sigma^2$ and
$X^4$, $X^5$ respectively.  
The modular integration in (\ref{delta}) can be done using the 
standard trick \cite{dkl}, where fundamental domain integrals
are unfolded to the strip (degenerated orbits ${\rm Det} M=0$) and to
the whole upper half plane (non-degenerated orbits ${\rm Det}M\neq 0$)
integrals of certain $SL(2,Z)$ representatives.
The final result is \cite{hm,ko}
\be
\Delta_{\rm gr}(T) = - 36 \log \left( T_2 \left|\eta (T)
\right|^4 \right) .
\label{deltaII}
\ee
Using the duality relations (\ref{dualitymap}) we can rewrite 
this in terms of the type I variables:
\bea
\Delta_{gr}(S) &=& - 36 \log \left( S_2 \left|\eta (S)
\right|^4 \right)
=- 36 \log S_2 + 12 \pi S_2 \nonumber\\ 
&&+ 72\sum_{N=1} \left(\sum_{M|N}\frac{1}{M}\right)
[e^{2\pi i N S}+e^{-2\pi i N \bar{S}}] .
\label{deltaI}
\eea
where $N|M$ stands for the partitions of $N$ ($N=LM$ with 
$N,L,M\in {\bf Z}$).
The first term in (\ref{deltaI}) corresponds to a logarithmic
divergence in the weak coupling limit 
$S_2=\frac{{\cal V}_6}{\lambda}\rightarrow\infty$, 
the second term
come from a disk diagram contribution and the rest is
an infinite sum of D-instanton corrections. 
The logarithmic divergent 
term in (\ref{deltaI}) is attributed to an IR divergence, but a 
complete understanding is still missing  
(see \cite{kiritsis,kp} for details). 

It has been shown in \cite{foerger} that ${\cal R}^2$ couplings 
receive a resulting non-vanishing one-loop contributions
from the Klein, annulus and Moebius in type I theory. The
absence of the one-loop term in the perturbative formula (\ref{deltaI})
suggests that the ${\cal R}^2$ result in type IIA should
correspond to a combination of ${\cal R}^2$ together
with other four derivative couplings
in the type I side. For instance, the authors 
of \cite{foerger} have shown that a suitable combination
of ${\cal R}^2$ and ${\cal F}_1^2{\cal F}_2^2$ can account for
the discrepancy. One can see that four-derivative couplings
involving the dilaton field will lead again to a 
one-loop expression similar to its closely related partner ${\cal R}^2$,
making them potential sources to take into account the 
perturbative discrepancy. 
Moreover, the duality relation
$G_{\mu\nu}^{IIA}={\cal V}_4 e^{-\phi}G_{\mu\nu}^{I}$ suggests that
in the translation of the type IIA ${\cal R}^2$ term in the type I 
variables four-derivative couplings constructed out of the dilaton 
and volume modulus should indeed be relevant.
Our instanton computation will not
give a definite answer to this question, but it will support these
latter sources. The comparison with the perturbative formula 
will account then only for the form of the instanton sums, leaving
an overall coefficient to be accounted for by the right combination
of four derivative terms in type I dual to the 
type IIA ${\cal R}^2$ coupling.

A similar formula (with $S$ replaced by $S_H$
in (\ref{deltaI})) describe the instanton corrections in 
the heterotic side. The
correct instanton action weights $e^{2\pi i N S_H}$ were reproduced
in \cite{kiritsis} from the classical action of heterotic
NS-fivebranes wrapped on the $T^6$ torus, however it is
hard to see how the determinant
factor $\sum_{N|M}\frac{1}{M}$ can be computed with our current
understanding of the NS-fivebrane physics
Fortunately, type I
spacetime instantons are associated to the more tractable
D-branes,  for which a CFT description is at our disposal
\cite{polchinski}. 

\subsection{D5-brane instanton corrections to ${\cal R}^2$-couplings} 

We are interested in computing the two-graviton correlation
function
\be
\left\langle V_g V_g\right\rangle_{\rm D5-inst}
\label{vgvg}
\ee
in the background of $N$ Euclidean D5 branes wrapping the $T^6$
spacetime torus. We take for the spacetime torus the limit  
in which the volume of a $T^4$ torus in $T^6$ becomes very small
($R_i\sim \sqrt{\alpha^\prime}$, $i=6,7,8,9$) keeping
$R_4,R_5$ fixed when the D5-brane theory decouples from
gravity $\alpha^\prime\rightarrow 0$. We will reduce the 
six-dimensional world-volume theory onto this two-dimensional plane
with light cone coordinates $X^{\pm}=X^4\pm X^5$.   

The classical part of the computation in the D5-instanton background
closely follows the one for the heterotic 
NS-fivebrane \cite{kiritsis} with 
obvious modifications. The low energy effective action describing N 
spinless Euclidean D5-branes wrapping once the $T^6$ spacetime 
torus can be written 
as \cite{tasi}
\bea
S_{\rm ND5}&=&N T_5\int d^6 \xi\left[e^{-\phi}\sqrt{\det\hat G}
+i {\tilde B}_{456789}\right]+\cdots=\nonumber\\
&&-2\pi i N S_I+
\frac{2\pi {\cal V}_4 U_2}{\lambda}
\int d^2 z h_{\mu\nu}
\sum_{t=1}^N D_z X_t^{\mu} D_{\bar{z}}X_t^{\nu}+...
\label{5b}
\eea 
where ${\cal V}_4$ is the volume of the small $T^4$ torus, $U$ the complex
structure of the large $T^2$-torus,   
$T_5=\frac{1}{(2\pi)^5{\alpha^\prime}^3}$ the fivebrane tension
\footnote{Notice that the $T_1,T_5$
tensions of type I D-branes are $\frac{1}{\sqrt{2}}$-the corresponding
tension in type IIB. The electric-magnetic quantization condition
implies then that D5 branes come always in pairs to account for the
extra factor of $\sqrt{2}$ \cite{gp}.   
$N$ here counts the number of type I D5-branes which
halves the number in the parent type IIB theory.}, 
$S_I$ is the complexified string coupling constant given by
(\ref{map}) and $h_{\mu\nu}$ the quantum fluctuation  
($G_{\mu\nu}=\eta_{\mu\nu}+h_{\mu\nu}$) 
around the flat metric, $\mu,\nu=0,1,2,3$. 
We choose the static gauge $X^a=\xi^a$
($a=4,5,..,9$) for the fivebrane coordinates, where the 
six-dimensional pullback metric
$\hat{G}$ is identified with the spacetime torus metric. 
The last term in (\ref{5b}) represents the coupling of the graviton
to the instanton background with covariant derivatives
\bea
D_{z} X_t^{\mu}&=&\partial_{z} X_t^{\mu}-\frac{1}{4}p_{\nu} 
S_t\gamma^{\nu\mu} S_t
\nonumber\\
D_{\bar{z}} X_t^{\mu}&=&\partial_{\bar{z}} X_t^{\mu}-\frac{1}{4}p_{\nu} 
\tilde{S}_t\gamma^{\nu\mu} \tilde{S}_t
\eea 
in the $(4,5)$ plane with complex
coordinates $z=\xi^4+U\xi^5$, $\partial_z=\frac{1}{U_2}(
\partial_{\xi_4}+U\partial_{\xi_5})$\footnote{Strictly speaking,
our analysis will be performed in the Minkowski
world-volume with time like coordinate $\xi^5$. Only at the end we
will come back to the Euclidean plane by an analytic continuation.}.
Quantum  
corrections around this background are described by an 
$Sp(N)$ gauge theory defined by the quantization 
of the massless modes of unoriented open strings ending on the D5 branes. 
    
The computation of the scattering amplitude (\ref{vgvg}) is very
similar to the perturbative computation in the one-loop worldsheet 
instanton background of the last section, with the worldsheet 
parameter $\tau$ replaced by the complex structure
$U$ of the target space $(4,5)$ torus. Graviton insertions can be expressed
as derivatives of the instanton action (\ref{5b}) with respect to
the metric fluctuations $h_{\mu\nu}$ $\mu=0,1,2,3$, bringing down
the $(4,4)$ needed fermionic zero modes $S_{0{\rm cm}}$, the right
power of momenta to reproduce the ${\cal R}^2$ kinematics and  
an overall $\frac{{\cal V}_4^2 U_2^2}{N^2 \lambda^2}$ factor. 
Unlike in references \cite{bgmn,ghmn}, we adopt 
the canonical normalization $S_{t}=\frac{1}{\sqrt{N}} S_{\rm cm}+....$
for the fermionic center of mass 
($\sum_{t=1}^N S_t \partial S_t=S_{\rm cm} \partial S_{\rm cm}+...$), 
which is responsible for the additional factor of $\frac{1}{N^2}$. 
The final result can then be written as
\bea
\Delta_{\rm gr}^{\rm inst}&=&\frac{{\cal V}^2_4U_2^2}{N^2\lambda^2}
\left\langle e^{-S_{\rm class}-S_{SYM}}
\right\rangle^\prime=\frac{{\cal V}^2_4}{\lambda^2}
\sum_N B^{ND5}_4 (e^{2\pi i N S}+e^{-2\pi i N \bar{S}})
\label{deltaD5}
\eea
in terms of the 
$\frac{1}{2}$-BPS index $B_4$ (\ref{b4}) of the $Sp(N)$   
gauge theory.
The prime in (\ref{deltaD5}) means that the trace does not
include the fermionic zero mode part already taken into account by
the vertex insertions, while the spacetime bosonic zero mode contribution
cancel the $U_2^2$ factor in the numerator. The overall $\frac{1}{N^2}$
factor in (\ref{deltaD5}) has been 
reabsorbed in $B_4^{\rm ND5}$ in order to make 
more transparent the comparison with the IIA perturbative result.  
Finally the $S_{class}=-2\pi i N S 
(2\pi i N \bar{S})$ represents the classical
(anti)instanton action.  

The rest of this section will be devoted to the computation of the 
$B^{ND5}_4$ index of the $Sp(N)$ gauge theory. 
For future reference, we will be slightly 
more general than what we really
need, by determining the whole BPS elliptic genus $\chi(v|\tau)$.  
We will follow the strategy of \cite{gmnt,bgmn}. The elliptic
genus, being invariant under any deformation of the gauge theory, in
particular under variation of the string coupling constant, can be
evaluated in the regime which is more convenient for our purposes.
We will compute it explicitly in the infrared limit, where
the theory will be argued to flow to an orbifold conformal fixed point. 

The low energy effective action associated to a system of
N parallel D5-branes in type I theory
can be obtained from the more familiar $U(2N)$   
gauge theory
describing $2N$ D5-branes in the parent type IIB theory. Type IIB
D5 brane fields $\Phi$ are projected onto $2N \times 2N$ 
matrices satisfying the 
$\Omega$-even simplectic condition \cite{rey}
\bea
 \Phi=\pm \Omega\Phi^T {\Omega}^{-1}~~~~~~~~~~~~~
\Omega=\sigma_2\times {\bf I}_N
\label{sp}
\eea
where $+(-)$ stands for the DD(NN) directions and $\sigma_i$ are
the Pauli matrices. In addition, anomaly cancellation requires the inclusion
of $32$ D9-branes and the corresponding open
string sectors.  
The resulting two-dimensional field content after dimensional
reduction on $T^4$ is defined
by all possible open strings ending on the D5-branes, and is given by\\
\\
\begin{tabular}{lllc}
Sector&   Bosons   &  Fermions & $ Sp(N)\times SO(32)$\\
5-5 $NN$& $A_{\alpha}$& & $N(2N+1),1$\\
5-5 $NN_I$& $a^{A^\prime \tilde{A}^\prime}$&
$\epsilon_-^{A \tilde{A}^\prime},
\epsilon_+^{A A^\prime}$ &$ N(2N+1),1$\\
5-5 $DD_E$& $ X^{A Y}$&
$\eta_-^{A^\prime Y},\eta_+^{\tilde{A}^\prime Y}$ & 
$N(2N-1),1$\\      
5-9 $ND$& $h^A$ & $\rho_-^{A^\prime}$,$\rho_+^{\tilde{A}^\prime}$ & 
$2N,32$ \\
\label{fc}
\end{tabular}\\
Here the subscript $E$ refers to the 4 directions transverse to the
D5-brane whereas $I$ refers to the 4 directions corresponding to the $T^4$
along which the dimensional reduction is performed. The corresponding
isometry groups $SO(4)_E$ and $SO(4)_I$ are decomposed as $SO(4)_E=
SU(2)_A\times SU(2)_Y$ and $SO(4)_I=SU(2)_{A^\prime}\times 
SU(2)_{\tilde{A}^\prime}$. $A_\alpha$ is a two-dimensional $Sp$
gauge field and $+(-)$ refer to left-(right-) moving fermions.
Fields in the same line are related by the ${\cal N}=(4,4)$ supersymmetry.
We will consider a generic type I background involving
Wilson lines on $T^4$, which break $SO(32)$ down to
$U(1)^{16}$.  In the strong coupling limit $g\rightarrow \infty$ 
of the Coulomb
branch of the above gauge theory, the  
off-diagonal fields get infinite masses and can be integrated out
(see \cite{gmnt} for a detailed analysis) leaving a free   
(up-to a Weyl group orbifolding)
conformal field theory in terms of the
Cartan components\footnote{We assume to be away from loci in the moduli space
where (5-9) fields can become massless due to cancellations between
$SO(32)$ and $Sp(N)$ Wilson lines \cite{witten2}.}
\begin{center}
$a^{A^\prime \tilde{A}^\prime}$,$\epsilon_-^{A \tilde{A}^\prime}$,
$\epsilon_+^{A A^\prime}$~~~~~$\sigma_3\times {\bf \lambda_N}$\\
$X^{A Y}$,$\eta_-^{A^\prime Y}$,
$\eta_+^{\tilde{A}^\prime Y}$~~~~${\bf I_2}\times {\bf \lambda_N}$\\
\end{center}
with ${\bf \lambda_N}$ an $N\times N$ matrix with diagonal entries.
The Weyl group of $Sp(N)$ is given by the semi-direct product 
$S_N\ltimes Z_2^N$, with $S_N$ permuting the $Z_2$'s factors, and $Z_2$'s 
acting as $\sigma_2$, and therefore reflecting the fields proportional to
$\sigma_3$ while leaving 
invariant the ones proportional
to the identity $\bf I_2$. The breaking of the $Sp(N)$ 
gauge group down to $S_N\ltimes Z_2^N$ can be understood in two steps. 
First by giving generic expectation values to the diagonal entries in $X$ 
(D5-brane positions) we
break the group to $Sp(1)^N$ with the Weyl group $S_N$ permuting
the branes. One can then further break each $Sp(1)$ to its
Weyl subgroup $Z_2$ by turning on the $SU(2)$ Wilson lines 
$a^{A^\prime \tilde{A}}$.
Alternatively, one can start from the type I theory and perform 
four T-dualites on the small $T^4$ directions. The $\Omega$- projection
goes under the T-duality map to $\Omega I_4$, introducing 16
5-orientifold planes whose charges are localy canceled with the inclusion of
D5-branes symmetrically distributed over the 16 fixed points. 
Carrying out the same steps as before on this effective 
$Sp(N)$ gauge theory, we are left with the N 
D-string sigma model moving on $({\bf R}^4\times T^4/Z_2)^N/S_N$ 

The resulting CFT can then be written in terms of a second quantized
string theory describing  
N copies of a type IIA string moving on the target space
\be
{\cal M}=\left[{\bf R}^4\times T^4/Z_2\right]^N/S_N
\label{cft}
\ee
The orbifold partition function is given by a sum 
over twisted sectors labeled by the conjugacy classes of the $S_N\ltimes Z_2^N$
orbifold group. In particular the sum over conjugacy classes in 
the permutation group $S_N$ runs over the decompositions 
\be
[g]=(1)^{N_1}(2)^{N_2}......(s)^{N_s}
\ee
with $\sum_s s N_s=N$. However, as it has been shown in 
\cite{gmnt}, only the sectors belonging to
the conjugacy classes of the kind $[g]=(L)^M$ (with $N=LM$)
will lead to a non trivial contribution to $\chi(v|\tau)$.
Sectors with strings of different lengths $[g]=(l_1)^{m_1}(l_2)^{m_2}...
$ with $l_1\neq l_2$ will contain always additional right moving 
fermionic zero modes leading to vanishing contributions to $\chi(v|\tau)$.     
The elliptic index $\chi(v|\tau)$ can then
be computed using the formula \cite{bgmn} for the N-symmetric product:
\footnote{Alternatively one can compute the 
helicity generating function $Z(v,\bar{v})$  
using the more general formula \cite{dvvm} and then
derive the BPS index from (\ref{b4}).}
\be
\chi_N(v|U)=
\frac{8}{N}\sum_{L,M}
\sum_{s=0}^{L-1}M^{-2}
\frac{\xi(Mv|\tilde{U})\theta_1^2(\frac{M v}{2}|\tilde{U})}
{\eta^6(\tilde{U})} 
\sum_{i=2,3,4} 
\frac{\theta_i^2(\frac{Mv}{2}|\tilde{U})}{\theta_i^2(0|\tilde{U})}
\label{genus}
\ee
with all modular functions evaluated in the induced 
space-time complex modulus $\tilde{U}=\frac{M U+s}{L}$.   
The relative $M^{-2}$ factor is determined from modular 
transformations from the untwisted sector \cite{bgmn,ghmn}, once
the overall $N^2$ factor brought down by the vertex insertions is
taken into account. Let us recall how these relative factors arise in the
present canonical normalization.
One start from $Tr_{untwisted}(L)^M$ trace in the untwisted
sector where one can umambiguously compute the partition function 
using the operator formalism. After a modular transformation 
$\tau\rightarrow -\frac{1}{\tau}$, to the
$g=(L)^M$-twisted sector we are left with the partition
function for $M$ copies of strings of length $L$ 
($q\rightarrow q^{\frac{1}{L}}$) weighted by an $L^{2M}$ factor. 
The projection by a $Z_M$ permutation elements remove the additional
fermionic zero modes, apart from the center of mass zero modes, leading
finaly to an $(\frac{L}{M})^2$ factor 
from the uncompact bosons and an $M^2$-factor from the right
moving fermionic $Z_M$-trace. After including the overall
$N^2$ factor we are left with the $M^{-2}$ result claimed above.

Substituting (\ref{genus}) in (\ref{b4}), one can see that oscillator
contributions from massive fermionic and bosonic modes cancel
against each other leaving the $\frac{1}{2}$-BPS index 
\bea
B_4^{ND5}&=&36\frac{1}{N}\sum_{N|L}L=36\sum_{N|M}\frac{1}{M}
\label{b4f}
\eea
Apart from a factor of two, the type I result (\ref{deltaD5}) 
in terms of this $B_4$ index reproduce exactly the instanton sum
in the formula (\ref{deltaI}). The extra factor 
$\frac{{\cal V}_4^2}{\lambda^2}$, coming from the coupling
of the metric to the instanton background (\ref{deltaD5}) 
is cancelled by a similar one coming from the spacetime measure
and the metrics used in the ${\cal R}^2$ contractions,
once the duality map 
$G_{\mu\nu}^{II}=\frac{{\cal V}_4 G_{\mu\nu}^{I}}{\lambda}$ 
is taken into account. 

One can easily extend the previous results to other four derivative 
couplings in the $D=4$ type I effective lagrangians. We can
consider for example ${\cal F}^4$ terms with ${\cal F}$ the $U(1)$
gauge fields coming from the reduction of the metric on the $T^2$ torus
($F_{\mu\nu}=\partial_{[\mu}G_{\nu ]i}$, with $i=4,5$).  
The analysis follows closely our previous one for the ${\cal R}^2$
computation, with an effective coupling of the gauge
field to the instanton background given in this case by
\be
S_{\rm ND5}=-2\pi i N S+\frac{2\pi {\cal V}_4}{\lambda}
\sum_{t=1}^N\left[(h_{\mu 5}+U h_{\mu 4})D_z X_t^\mu+
(h_{\mu 5}-\bar{U} h_{\mu 4})D_{\bar{z}} X_t^\mu\right]+...
\label{sclassf}
\ee
The four vertex insertions provide then the $(4,4)$ fermionic zero modes
needed to get a non trivial answer and the four powers of momenta
to reproduce the ${\cal F}^4$ kinematics. In addition each ${\cal F}_4$
$({\cal F}_5)$ insertion carry an additional 
$\frac{{\cal V}_4}{\lambda}$($\frac{{\cal V}_4 U_2}{\lambda}$)
factor from (\ref{sclassf}) (for simplicity 
we take a rectangular spacetime torus, i.e. $U=iU_2$). 
The result reproduces exactly the 
contributions from non-degenerate orbits to the one-loop
formula for similar ${\cal F}^4$ terms in type IIA. In particular, 
we can see that the N-instanton contribution to ${\cal F}_4^2{\cal F}_5^2$
is $N^2$ times the previously found for ${\cal R}^2$ (\ref{deltaD5}).
Therefore the mixing of ${\cal R}^2$ and ${\cal F}_4^2{\cal F}_5^2$
cannot account for the absence of the
one-loop term  in (\ref{deltaI}), without destroying the agreement at the 
non-perturbative level. 
On the other hand, one can easily see that four derivative
couplings involving the dilaton, mentioned in section (2.1),  have  the same
$N$ dependence of the ${\cal R}^2$ term, making them potential 
candidates to account for the perturbative discrepancy and the factor
of 2 in the non-perturbative contribution.

Finally, we would like to stress that the result (\ref{genus}) 
is stronger than what we really need. Indeed, as we have discussed,
${\cal R}^2$ couplings in type I theory receive
contributions only from the $\frac{1}{2}$-BPS states (\ref{b4f}).
However, one can see that even the right $\frac{1}{4}$-BPS degeneracies,
are reproduced by the CFT elliptic genus (\ref{genus}).
Indeed according to the type IIA/type I duality map, 
a fundamental type IIA string with winding $N$ and momentum
$k$ is mapped in the type I theory to a bound state of 
N D5 branes and k Kaluza-Klein momenta. 
The masses of the two objects agree according to (\ref{map}). 
Multipicities and charges for the bound states can be read from the 
longest string sector in the CFT elliptic
genus (the only orbifold sector representing a true one-particle 
state as shown in \cite{gmnt}). 
Specifying to this sector, $L=N,M=1$ in 
(\ref{genus}), we are left with
\be
\chi_N(v|\tau)=
\frac{8}{N}
\sum_{s=0}^{N-1}
\frac{\xi(v)\theta_1^2(\frac{v}{2})}
{\eta^6} 
\sum_{i=2,3,4} 
\frac{\theta_i^2(\frac{v}{2})}{\theta_i^2(0)}(q^{\frac{1}{N}}
e^{-\frac{2\pi i s}{N}})
\label{bpsg}
\ee
The sum over $s$ projects onto states satisfying 
\bea
k\equiv \frac{N_L-c_L}{N}\in {\bf Z}\label{lm}
\eea  
which reproduce the level matching condition in the fundamental 
side (\ref{zvvII}) after putting right moving modes on their ground
state ($N_R-c_R=0$) and identifying $k$ with the KK momentum
along the direction where the string wraps. 
The bound state degeneracies are defined by the coefficients in the
expansion of (\ref{bpsg}) in powers of $q^{\frac{N_L-c_L}{N}}$.
In particular the ground state multiplicities ($q^0$ order) 
counts the number of ultrashort supermultiplets, associated
to bound states of N D5-branes, while degeneracies in the excited 
left moving part of the CFT (coefficient of $q^k$ with $k>0$)
are associated to bound states of N D5 branes with k units of 
Kaluza-Klein longitudinal momenta, which sit in intermediate
supermultiplets.    
The expansion clearly coincide with the similar one for the 
fundamental result (\ref{zvvII}). 
We conclude then that the whole spectrum of masses, charges and 
multiplicities of $\frac{1}{2}$- and $\frac{1}{4}$-BPS excitations
in the D5-brane system agree with the S-duality predictions.
In the next section we will give an application of these results.

\section{ $\frac{1}{4}$-BPS saturated couplings in type I theory}
 
In this section we consider higher derivative couplings
which are sensitive to $\frac{1}{4}$-BPS contributions. 
Instanton corrections to type I thresholds will translate
again into an infinite sum of one-loop contributions coming
from $T^2$-wrapping modes of type IIA fundamental strings running
in the loop, the novelty being in the fact that now 
$\frac{1}{4}$-BPS fundamental strings will be the relevant ones.
The type IIA perturbative computation follows with slight
modifications from the ones appering in \cite{ls}, reference 
that can be consulted
for details and a more complete discussion. 
We will consider in particular the (4+g)-derivative couplings 
$\partial{\cal F}_L\partial\bar{{\cal F}}_L {\cal F}_L^g$,
with 
${\cal F}_{\mu\nu\, L}\equiv\partial_{[\mu}(G_{\nu]z}+B_{\nu]z})$ the
left moving combination of $U(1)$ gauge fields arising from the
reduction of the metric and antisymmetric tensor on $T^2$

The relevant string amplitude is given by
\be
{\cal A}_{g}=\left\langle V_{L}^{g+1} V_{\bar{L}} \right\rangle
\label{ajk}
\ee
with vertex operators
\bea
V_L&=&(G_{\mu z}+B_{\mu z})\int d^2 z
\left(\partial X^\mu -\frac{1}{4}p_{\rho}S 
\gamma^{\mu\rho} S\right)
\left(\bar{\partial} Z -\frac{1}{4}p_{\sigma}\tilde{S} 
\gamma^{z\sigma}\tilde{S}\right)e^{ipX}\nonumber\\
V_{\bar{L}}&=&(G_{\mu\bar{z}}+B_{\mu\bar{z}})\int d^2 z
\left(\partial X^\mu -\frac{1}{4}p_{\rho}S 
\gamma^{\mu\rho} S\right)
\left(\bar{\partial}\bar{Z} -\frac{1}{4}p_{\sigma}\tilde{S} 
\gamma^{\bar{z}\sigma}\tilde{S}\right)e^{ipX}.
\label{vll}
\eea
At the order of momenta, we are interested in, one can see that 
the leftt moving part in the vertices (\ref{vll}) enters only
through their zero mode part. Indeed, after soaking the $(4,4)$
fermionic zero modes the remaining extra $g$ powers of momenta
are necesarily carried out by the right moving pieces of the vertices
(\ref{vll}) and therefore 
the left moving part reduces effectively
to the $p_L,\bar{p}_L$ bosonic zero modes of $\bar{\partial} Z$ 
and $\bar{\partial}\bar{Z}$ respectively. The right moving part
can then be replaced by the first order in the momentum effective  
vertex
\be
V_{\rm eff}={\cal F}_{\mu\nu\, L,\bar{L}} \int d^2z\,
(X^\mu \partial X^\nu-\frac{1}{4}S\gamma^{\mu\nu}S)
\label{veff}
\ee
The ${\cal A}_{g}$ string amplitudes reduce then to a correlation 
function of $g$ effective vertices (\ref{veff}), which 
exponentiate to \cite{agnt,ms}
\bea
{\cal A}_{g}&=&\partial{\cal F}_L\partial\bar{{\cal F}}_L 
\left\langle p_L^g \tau_2^g\, 
\frac{\partial^{g}}{\partial v^{g}}
\left[e^{-S_{\rm free}+\frac{v}{\tau_2} V_{\rm eff}}\right]  
\right\rangle_{v=0}^\prime\nonumber\\
&&=\partial{\cal F}_L\partial\bar{{\cal F}}_L
{\cal F}_L^g
\int \frac{d^2 \tau}{\tau_2}\tau_2^g\sum_{(p_L,p_R)}  
q^{\frac{1}{2}|p_L|^2}\bar{q}^{\frac{1}{2}|p_R|^2} p_L^g\, 
f_{g}(\tau)
\label{amp}
\eea
with $f_{g}(\tau)$ holomorphic indices generated by the 
oscilator contribution $\chi_{\rm osc}(v|\tau)$ to the type IIA 
elliptic genus $\chi(v|\tau)=\chi_{\rm osc}(v|\tau)\Gamma_{2,2}$ through
\be
f_{g}(\tau)\equiv \frac{\partial^{g}}{\partial v^{g}}
\chi_{\rm osc}(v|\tau)|_{v=0}
\label{fjk}
\ee
We have denoted again by prime, the omission of the fermionic
zero mode trace in (\ref{amp}). 
We can now follow the results of \cite{foerger,ls}, where the 
modular integral in (\ref{amp}) has been performed
for an arbitrary holomorphic function $f_g(\tau)$.
The result can be written as
\bea
{\cal A}_{g}&=&\partial{\cal F}_L\partial\bar{{\cal F}}_L {\cal F}_L^g
\left(\frac{i}{\pi}\sqrt{\frac{2U_2}{T_2}}\right)^g\sum_{m_1,n_2,n}
d(n)\frac{(T_2 U_2)^g}{\pi^g(m_1 T_2+\frac{n}{m_1}U_1)^{2g}}
\frac{\partial^g}{\partial\alpha^g}\frac{\partial^g}{\partial\nu^g} 
I(\alpha,\nu)
|_{\nu=0,\alpha=1}\nonumber
\eea
with $d(n)$ the coefficients in the $q$-expansion $f_g=\sum_n d(n) q^n$
and 
\bea
I(\alpha,\nu)&=&\frac{2}{\sqrt{b}}e^{-2\pi\sqrt{\alpha b}
(m_1 T_2+\frac{n}{m_1} U_2)}e^{-2\pi i \phi}\nonumber\\
\sqrt{b}&=&|n_2+\frac{\nu}{4 U_2}|\nonumber\\
\phi&=&n_2 m_1 T_1-n\frac{n_2}{m_1}U_1-i\nu(\frac{n}{m_1}+m_1\frac{T_2}{U_2})
\eea
The $T^{II}_2$ expansion of the above formula translate into a series of
perturbative corrections (subleading orders in $S^I_2$) around the 
instanton background. We will consider in the following only the
leading order in this expansion. The analysis of subleading 
quantum corrections to this result can be done following the techniques
in \cite{ghmn}. At this order $\nu$ derivatives
hit always the terms proportional to $\nu T_2$ and set
the remaining $\nu$-dependent terms to zero. The $\alpha$ derivatives
hit all the time the exponential leading to an overall 
$(\pi n_2 m_1 T_2)^g$ factor. Altogether we are left with 
\bea
{\cal A}_{g}&=&\partial{\cal F}_L\partial\bar{{\cal F}}_L {\cal F}_L^g
\left(\frac{i}{\pi}\sqrt{\frac{2U_2}{T_2}}\right)^g\sum_{m_1,n_1,n_2} 
\frac{e^{2\pi i n_2 m_1 T}}
{n_2 m_1}\,n_2^g f_g \left( \frac{n_2 U+n_1}{m_1}\right)+h.c+\,\dots\nonumber\\
&=&\partial{\cal F}_L\partial\bar{{\cal F}}_L {\cal F}_L^g    
\frac{e^{2\pi i n_2 m_1 T}}
{n_2 m_1}\left.\frac{\partial^g}{\partial v^g}\chi\left(\frac{i\sqrt{2U_2}}
{\pi\sqrt{T_2}}
\, n_2 v|\frac{n_2 U+n_1}{m_1}\right)\right|_{v=0}+h.c+\,\dots
\label{fund}
\eea
where h.c. stands for hermitian conjugation and dots for the higher
orders in $T_2$ expansion. 
It is now straight to compare this result with the corresponding
string amplitudes in the instanton background. Indeed, the
partition function for the ND5-instanton in the presence of 
a background (\ref{veff}) is described by the elliptic
genus (\ref{genus}), whose $v$-derivatives coincide with the
fundamental result (\ref{fund}) after trivial identifications.

The presence of perturbative corrections around the instanton 
background is a new feature of 
these higher derivative couplings, to be contrasted with 
the ${\cal R}^2$ case. It would be interesting to compare (along
the lines of \cite{ghmn}) the fundamental and D-instanton results 
for these quantum corrections,
where the Born-Infeld nature of the instanton couplings are strongly
tested. 
Notice in particular that insertions of $F_{\bar{L}}$ gauge fields
appears already at a quantum level. A complete analysis of
the quantum subleading terms would determine the moduli dependence 
of the additional
couplings in the effective lagrangian.

Notice also that for the case $g=4$, the result
(\ref{fjk}) is proportional to $\Gamma_{2,2}$. 
Indeed $f_4\Gamma_{2,2}$ is nothing but the
helicity supertrace $B_6\equiv {\rm Str}\lambda^6$ for type IIA
string theory on $K3\times T^2$, where $\lambda$ is the
4-dimensional physical helicity. $B_6$ in this case is therefore
only sensitive to short multiplets \cite{kirt}, the intermediate ones
give vanishing contribution to this helicity supertrace,
however this accident will clearly not persist for the rest of the
asymmetric supertraces generated by $\chi(v|\tau)$.
%even though it can be verified that, in general, they do not pair up into
%long multiplets.

\section{Conclusions}

In this paper we have considered instanton corrections to four and 
higher derivative couplings in $D=4$ type I 
string vacua with sixteen supercharges. 
We restricted our attention to couplings for which Euclidean 
D5-brane wrapping the $T^6$ torus represent the only source
of instanton corrections.
D-string instantons have been extensively studied
in \cite{bfkov}-\cite{ghmn} and are fairly better understood. 
The couplings we consider  are also
special in the sense that they are sentitive only to states
sitting in short and intermediate multiplets of the
$N=4$ supersymmetry. 

We have worked it out the details of the 
instanton sums for ${\cal R}^2$ thresholds in 
toroidal compactifications of type I string theory.
The instanton sums translate, under the duality map,
into a sum over wrapping modes
of fundamental type IIA strings on the $T^2$ part
of $K_3\times T^2$.  
We argued that the relevant 6-dimensional $Sp(N)$ gauge theory
flows in the infrared to an orbifold conformal fixed point, after
dimensional reduction to 1+1 dimensions. The elliptic
genus for the orbifold CFT was computed in this limit and
shown to reproduce correctly the 
whole spectrum of $\frac{1}{2}$ BPS masses, charges and 
multiplicities, as required by type I/type IIA duality.
As a consequence, the whole infinite sum of instanton corrections
to four derivative couplings agree with the expected result
from the IIA fundamental string side. 

The proposed CFT 
reproduce also the right multiplicities for 
$\frac{1}{4}$-BPS states associated to bound states of D5 branes
and KK momenta in the type I theory, providing several more examples 
of higher derivative couplings where
the worldsheet/spacetime instanton correspondence works properly.

Besides supporting type IIA/type I duality in $D=4$ dimensions,
we believe the results reported here motivate a deeper study
of D5-brane conformal field theory description in type I theory, rather
less understood than its type IIB parents. 
An exciting direction would be the study of conformal field theory
description of other $\frac{1}{4}$-BPS magnetic charges in the
type I side. In particular, unlike in the type IIB case where
an available orbifold CFT description of BPS excitations in the D1-D5 
system exists, the type I analog is still missing. The role
of these states in the ADS/CFT correpondence, as well as of
the $\frac{1}{4}$-BPS bound states studied here, deserves
a deeper attention.

\vskip 0.5in
{\bf Acknowledgements}

We are particularly grateful to E. Gava and K.S. Narain for several
discussions and comments. We thank E. Kiritsis for suggestions and 
M. Bianchi for helpful discussions.

\rnc{\Large}{\normalsize}

\end{document}